\pdfoutput=1

\RequirePackage{fixltx2e}
\RequirePackage{fix-cm}

\documentclass[%
floatfix,
showkeys,
twocolumn, %
nofootinbib, %
superscriptaddress, %
]{revtex4-1}

\usepackage{cmap}

\usepackage{ucs}
\usepackage[utf8x]{inputenc}
\usepackage[T1,T2A]{fontenc}
\usepackage[german,russian,english]{babel}

\usepackage[sort&compress]{natbib}
\usepackage{amsmath}
\usepackage{amssymb}

\usepackage{mathtools}
\mathtoolsset{
showonlyrefs,
mathic = true
}

\allowdisplaybreaks

\usepackage{hyperref}
\hypersetup{backref,
 colorlinks=false}
\hypersetup{pdfborder=0 0 0}

\usepackage{microtype}
\UseMicrotypeSet[protrusion]{alltext}

\usepackage{graphicx}

\usepackage[scanall]{psfrag}

\usepackage{listings}
\usepackage{listingsutf8}
\usepackage{fixltx2e}

\usepackage{nicefrac}

\makeatletter
\def\ps@pprintTitle{%
     \let\@oddhead\@empty
     \let\@evenhead\@empty
     \let\@oddfoot\@empty
     \let\@evenfoot\@oddfoot}
\makeatother

\usepackage{physics}
\usepackage{tensor}

\usepackage{float}

\usepackage{algorithm2e}

\newcommand{\crd}[1]{\underline{\vphantom{j}{#1}}\,}

\newcommand{\setR}{\mathbb{R}}

\newcommand{\setN}{\mathbb{N}}

\begin{document}
\graphicspath{{image/}}

\title{Implementing a Method for Stochastization of One-Step Processes in a Computer Algebra System}

\author{M. N. Gevorkyan}
\email{gevorkyan_mn@rudn.university}
\affiliation{Department of Applied Probability and Informatics,\\
  Peoples' Friendship University of Russia (RUDN University),\\
  6 Miklukho-Maklaya str., Moscow, 117198, Russia}

\author{A. V. Demidova}
\email{demidova_av@rudn.university}
\affiliation{Department of Applied Probability and Informatics,\\
  Peoples' Friendship University of Russia (RUDN University),\\
  6 Miklukho-Maklaya str., Moscow, 117198, Russia}

\author{T. R. Velieva}
\email{trvelieva@gmail.com}
\affiliation{Department of Applied Probability and Informatics,\\
  Peoples' Friendship University of Russia (RUDN University),\\
  6 Miklukho-Maklaya str., Moscow, 117198, Russia}

\author{A. V. Korolkova}
\email{korolkova_av@rudn.university}
\affiliation{Department of Applied Probability and Informatics,\\
  Peoples' Friendship University of Russia (RUDN University),\\
  6 Miklukho-Maklaya str., Moscow, 117198, Russia}

\author{D. S. Kulyabov}
\email{kulyabov_ds@rudn.university}
\affiliation{Department of Applied Probability and Informatics,\\
  Peoples' Friendship University of Russia (RUDN University),\\
  6 Miklukho-Maklaya str., Moscow, 117198, Russia}
\affiliation{Laboratory of Information Technologies\\
  Joint Institute for Nuclear Research\\
  6 Joliot-Curie, Dubna, Moscow region, 141980, Russia}

\author{L. A. Sevastianov}
\email{sevastianov_la@rudn.university}
\affiliation{Department of Applied Probability and Informatics,\\
  Peoples' Friendship University of Russia (RUDN University),\\
  6 Miklukho-Maklaya str., Moscow, 117198, Russia}
\affiliation{Bogoliubov Laboratory of Theoretical Physics\\
  Joint Institute for Nuclear Research\\
  6 Joliot-Curie, Dubna, Moscow region, 141980, Russia}

\begin{abstract}
  When modeling such phenomena as population dynamics, controllable
  flows, etc., a problem arises of adapting the existing models to a
  phenomenon under study. For this purpose, we propose to derive new
  models from the first principles by stochastization of one-step
  processes. Research can be represented as an iterative process that
  consists in obtaining a model and its further refinement. The number
  of such iterations can be extremely large. This work is aimed at
  software implementation (by means of computer algebra) of a method
  for stochastization of one-step processes. As a basis of the
  software implementation, we use the SymPy computer algebra
  system. Based on a developed algorithm, we derive stochastic
  differential equations and their interaction schemes. The operation
  of the program is demonstrated on the Verhulst and Lotka--Volterra
  models.
\end{abstract}

  % \keywords{traffic active management, control theory,
  %   self-oscillating mode}

\maketitle

\section{Introduction}

  Many physical and technical phenomena can be described in the
  framework of the statistical approach. Usually, a model is selected
  that is sufficiently complete to reflect a phenomenon under study,
  and then this model is refined. A question arises as to how this
  refinement should be carried out, because this process is quite
  ambiguous. Models implement certain first principles. We believe
  that, if a model is constructed based on the first principles, then
  introduced modifications will express the internal structure of the
  model. To describe phenomena under study (data communication
  networks, control systems, and population dynamics), we use the
  model of one-step processes~\cite{gardiner:stochastic::en,van-kampen:stochastic::en}..

  We developed a stochastization
  technique that, based on the first principles, yields a stochastic
  model and a deterministic model corresponding to it~\cite{kulyabov:2016:ecms:one-step,kulyabov:2016:dccn-springer:diagram,kulyabov:2015:conf:mmcp-2015:doi,kulaybov:2017:ecms:wind,kulyabov:2016:rk-stochastic}.
  Research
  process is iterative: from a deterministic model, we obtain a
  primary model, from which we obtain a stochastic model that
  correlates with a deterministic model, and based on this
  correlation, we obtain a refined primary model. Then, the process
  repeats.

  Stochastization formalism can be implemented in different
  ways. Currently, we use a state vectors representation
  (combinatorial approach) and an occupation numbers representation
  (operator approach)~\cite{grassberger:1980:fock-space, tauber:2005,
    janssen:2004, mobilia:2005}.

  In the case of the combinatorial
  approach, all operations are carried out in the state-vector space
  of a system. During all model manipulations, we deal with a
  particular system. As a result, we obtain a description in the form
  of a differential equation. This approach is convenient when
  designing a model because it enables comparison with other models.

  In the operator approach, we abstract ourselves from a particular
  implementation of a system under study and deal with abstract
  operators. The space of state vectors is used only at the end of
  computations. Moreover, a particular operator algebra is selected
  based on the symmetry of a problem. This approach is convenient for
  theoretical constructions.

  Often, it is required to find a
  stochastic model equivalent to a previously designed deterministic
  model. For this purpose, we use the combinatorial approach. In this
  representation, the stochastic model has the form of a differential
  equation, which facilitates its comparison with the original model.

This paper is organized as follows. Section~\ref{sec:notation} introduces basic
notations and conventions. A method for stochastization of one-step
processes, as well as its components, is described in Section~\ref{sec:method}.
Section~\ref{sec:cas} describes a software complex that implements the
stochastization method. In Subsection~\ref{sec:cas:compare}, we compare some computer
algebra systems to select a basis for the software
implementation. Subsection~\ref{sec:programm} considers the main code fragments of
the stochastization program. Practical application of the method is
demonstrated in Subsections~\ref{sec:model:verhulst} and~\ref{sec:pp-model}.

\section{Notations and conventions}
\label{sec:notation}

\begin{enumerate}
\item  For tensors, we use the abstract index notation~\cite{penrose-rindler:spinors::en}, i.e., a
  tensor as a whole is denoted by a simple superscript (e.g., $x^{i}$), while
  tensor components are denoted by underlined superscripts (e.g., $x^{\crd{i}}$).
  
\item  We adhere to the following conventions. Latin indices from the
  middle of the alphabet ($i$, $j$, and $k$) are associated with the
  state-vector space of a system. Latin indices from the beginning of
  the alphabet ($a$) are associated with the Wiener-process space. Greek
  indices ($\alpha$) denote a number of different interactions in kinetic
  equations.
\end{enumerate}

\section{General review of the methodology}
\label{sec:method}

We developed an empirical technique for stochastization of one-step
processes. It is formalized in such a way that, to use this technique,
it is sufficient that the initial problem be represented as a Markov
process. However, only some of its steps are explicitly
algorithmized.

\begin{figure}
  \centering
  \includegraphics[width=\linewidth]{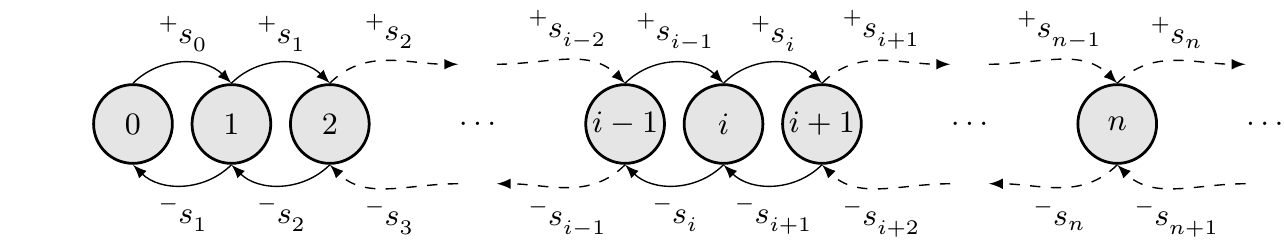}
  \caption{One-step process}
  \label{fig:one-step_process}
\end{figure}

At the first step, the model is reduced to a one-step process (see
Fig.~\ref{fig:one-step_process}). Then, this process needs to be represented in the form of
interaction
schemes~\cite{kulyabov:2014:icumt-2014:p2p,kulyabov:2015:conf:mmcp-2015:doi}. Analogs
of interaction schemes are
equations of chemical kinetics, particle response, etc.

For interaction schemes, we developed an algorithm (more precisely,
a family of algorithms where different algorithms correspond to
different approaches) that derives the master equation from
interaction schemes. This equation~\cite{van-kampen:stochastic::en,
  gardiner:stochastic::en}, however, usually has a
fairly complex structure, which hinders its solution and
investigation. At the next step, we obtain approximate models in the
form of the Fokker–Planck and Langevin equations.

The proposed approach assumes that research is an iterative process:
approximate models are refined, leading to the refinement of the
original interaction schemes.

To construct models, we use the following iterative algorithm (algorithm~\ref{alg:stochast:real}).   

\begin{algorithm}
  \caption{Algorithm of research}
  \label{alg:stochast:real}
  \SetAlgoLined
  \KwData{deterministic equation}
  \KwResult{stochastic equation}
  \While{correspondence between equations isn't established}{
    interaction scheme
    $\leftarrow$
    deterministic equation \;
    stochastic equation
    $\leftarrow$
    interaction scheme\;
    \eIf{stochastic equation satisfies the model}{
      terminate stochastization\;
    }{
      reiterate\;
    }
  }
\end{algorithm}

Unfortunately, the most part of this algorithm is not
formalized. Derivation of interaction schemes and determination of
correspondence between equations are quite difficult to formalize. And
yet, we managed to formalize the derivation of stochastic equations
from interaction schemes (see algorithms~\ref{alg:stochast:full} and~\ref{alg:stochast:simple}).

\begin{algorithm}
  \SetAlgoLined
  \KwData{interaction scheme~\eqref{eq:chemkin}}
  \KwResult{Langevin equation~\eqref{eq:langevin} and~\eqref{eq:k-langevin}}
  \Begin{%
    \emph{System state operators}\\
    $I_j^{i \alpha}$, $F_j^{i \alpha}$  $\leftarrow$
    interaction scheme
    $\eqref{eq:chemkin}$\;
    \emph{State transition operator}\\
    $r_j^{i \alpha}$ $\leftarrow$
    $I_j^{i \alpha}$, $F_j^{i \alpha}$ $\eqref{eq:r_i}$\;
    \emph{Transition rates}\\
    $\tensor*[^{+}]{s}{_{\crd{\alpha}}}$,
    $\tensor*[^{-}]{s}{_{\crd{\alpha}}}$ $\leftarrow$
    $I_j^{i \alpha}$, $F_j^{i \alpha}$, $r_j^{i \alpha}$ $\eqref{eq:s-pm}$\;
    master equation
    $\leftarrow$
    $\tensor*[^{+}]{s}{_{\crd{\alpha}}}$,
    $\tensor*[^{-}]{s}{_{\crd{\alpha}}}$,
    $r_j^{i \alpha}$
    $\eqref{eq:one-step:master:full}$ \;
    Fokker–Planck equation
    $\leftarrow$
    master equation
    $\eqref{eq:kramers-moyal}$, $\eqref{eq:fp_nD}$ \;
    Langevin equation
    $\leftarrow$
    Fokker--Planck equation
    $\eqref{eq:langevin}$, $\eqref{eq:k-langevin}$ \;
  }
  \caption{Stochastization algorithm}
  \label{alg:stochast:full}
\end{algorithm}

\begin{algorithm}
  \SetAlgoLined
  \KwData{interaction scheme~\eqref{eq:chemkin}}
  \KwResult{Langevin equation~\eqref{eq:langevin} and~\eqref{eq:k-langevin}}
  \Begin{%
    \emph{System state operators}\\
    $I_j^{i \alpha}$, $F_j^{i \alpha}$  $\leftarrow$
    interaction scheme
    $\eqref{eq:chemkin}$\;
    \emph{State transition operator}\\
    $r_j^{i \alpha}$ $\leftarrow$
    $I_j^{i \alpha}$, $F_j^{i \alpha}$ $\eqref{eq:r_i}$\;
    \emph{Transition rates}\\
    $\tensor*[^{+}_{\text{fp}}]{s}{_{\crd{\alpha}}}$,
    $\tensor*[^{-}_{\text{fp}}]{s}{_{\crd{\alpha}}}$ $\leftarrow$
    $I_j^{i \alpha}$, $F_j^{i \alpha}$, $r_j^{i \alpha}$ $\eqref{eq:s-pm:exp}$\;
    Langevin equation
    $\leftarrow$
    $\tensor*[^{+}_{\text{fp}}]{s}{_{\crd{\alpha}}}$,
    $\tensor*[^{-}_{\text{fp}}]{s}{_{\crd{\alpha}}}$,
    $r_j^{i \alpha}$
    $\eqref{eq:langevin}$, $\eqref{eq:k-langevin}$ \;
  }
  \caption{Simplified stochastization algorithm}
  \label{alg:stochast:simple}
\end{algorithm}

This process is not complex but it is rather time-consuming. Below, we
describe its main stages in more details.

\subsection{Interaction Schemes}
\label{sec:schema}

The state of a system is described by a state vector $\varphi^{i} \in \setR^n$, where $n$ is the
dimension of the system. The operators $I^{i}_{j} \in \setN^{n}_{0}
\times \setN^{n}_{0}$ and
$F^{i}_{j} \in \setN^{n}_{0} \times \setN^{n}_{0}$
define the system state
before and after interaction, respectively. The component indices of
the system dimension take the values $\crd{i},\crd{j} =
\overline{1,n}$. As a result of interaction, the
system transits from one state to another.

There are $s$  types of different interactions in the system. Hence,
instead of the operators $I^{i}_{j}$ and $F^{i}_{j}$, we consider the
operators
$I^{i \alpha}_{j} \in \setN^{n}_{0} \times \setN^{n}_{0} \times
\setN^{s}_{+}$
and
$F^{i \alpha}_{j} \in \setN^{n}_{0} \times \setN^{n}_{0} \times
\setN^{s}_{+}$. The
component indices for the number of interactions take the values $\crd{\alpha} = \overline{1,s}$.

nteraction among system elements is described by interaction schemes.
\begin{equation}
  \label{eq:chemkin}
  I^{i \crd{\alpha}}_{j} \varphi^j
  \xrightleftharpoons[{\tensor*[^{-}]{k}{_{\crd{\alpha}}}}]{\tensor*[^{+}]{k}{_{\crd{\alpha}}}}
  F^{i \crd{\alpha}}_{j} \varphi^{j},
  \qquad \crd{\alpha} = \overline{1,s}.
\end{equation}

Here, Greek indices specify the number of interactions and Latin
indices denote the dimension of the system. The coefficients
$\tensor*[^{+}]{k}{_{\crd{\alpha}}}$
and
$\tensor*[^{-}]{k}{_{\crd{\alpha}}}$
represent the intensity (rate) of interaction.

The state of the system is given by the operator
\begin{equation}
  \label{eq:r_i}
  r_j^{i \crd{\alpha}} = F_j^{i \crd{\alpha}} -I_j^{i \crd{\alpha}}.
\end{equation}

Thus, one interaction step $\crd{\alpha}$ in the forward and backward
directions can be represented (respectively) as
\begin{equation}
  \begin{gathered}
    \varphi^{i}  \rightarrow \varphi^i + r^{i \crd{\alpha}}_{j} \varphi^{j},\\
    \varphi^{i} \rightarrow \varphi^{i} - r^{i \crd{\alpha}}_{j} \varphi^{j}.
  \end{gathered}
\end{equation}

Most often, the model is constructed in such a way that the tensors
$I^{i \crd{\alpha}}_{j}$ and $F^{i \crd{\alpha}}_{j}$
are diagonal in terms of Latin indices. Hence, we can explicitly
use the diagonal sections of these matrices and rewrite~\eqref{eq:chemkin}  in a more
common form (as a system of linear equations)~\cite{van-kampen:stochastic::en}:
\begin{equation}
  \label{eq:chemkin2}
  I^{i \crd{\alpha}}_{j} \varphi^j \delta_i
  \xrightleftharpoons[{\tensor*[^{-}]{k}{_{\crd{\alpha}}}}]{\tensor*[^{+}]{k}{_{\crd{\alpha}}}}
  F^{i \crd{\alpha}}_{j} \varphi^{j} \delta_i,
\end{equation}
where $\delta_{\crd{i}} = (1,\ldots,1)$.

We also use the following designations:  
\begin{equation}
  \label{eq:n^i-notion}
  I^{i \crd{\alpha}} := I^{i \crd{\alpha}}_{j} \delta^{j}, 
  \quad F^{i \crd{\alpha}} := F^{i \crd{\alpha}}_{j} \delta^{j}, \quad
  r^{i \crd{\alpha}} := r^{i \crd{\alpha}}_{j} \delta^{j}.
\end{equation}

\subsection{Master Equation}
\label{sec:master-equation}

For one-step processes, as a kinetic equation, we consider the master equation~\cite{van-kampen:stochastic::en,
  gardiner:stochastic::en}. 
\begin{multline}
  \label{eq:master:trans}
  \pdv{p(\varphi_{2},t_{2}|\varphi_{1},t_{1})}{t} = \int \bigl[
  w(\varphi_{2}|\psi,t_{2}) p(\psi,t_{2}|\varphi_{1},t_{1}) 
  - {} \\ {} -
  w(\psi|\varphi_{2},t_{2}) p(\varphi_{2},t_{2}|\varphi_{1},t_{1}) 
  \bigr] \dd{\psi},
\end{multline}
where $w(\varphi|\psi,t)$ is the probability of transition from state
$\psi$ to state $\varphi$ in a unit time. 

The master equation can be regarded as an implementation of the
Kolmogorov equation. However, the master equation is more convenient
and has a direct physical interpretation~\cite{van-kampen:stochastic::en}.

Having fixed the initial values $\varphi_{1},t_{1}$, we can write this equation for a subensemble: 
\begin{equation}
  \label{eq:master:subansemble}
  \pdv{p(\varphi,t)}{t} = \int
  \qty[
  w(\varphi|\psi,t) p(\psi,t) -
  w(\psi|\varphi,t) p(\varphi,t)
  ] \dd{\psi}.
\end{equation}

If a  domain of $\varphi$ is a discrete one then
the~\eqref{eq:master:subansemble} can be written as follows
(the states are numbered by $n$ and $m$):
\begin{equation} 
  \label{eq:mas_eq}
    \pdv{p_{n}(t)}{t} = \sum\limits_{m} 
    \qty[w_{nm} p_{m}(t) - w_{mn} p_{n}(t)],
\end{equation}
where the $p_{n}$ is the probability of the system to be in a state $n$ at time $t$,
$w_{nm}$ is the probability of transition from the state $m$ into the state $n$ per unit time.

For a system described by one-step processes, there are two types of
transitions from one state to another that occur as a result of
interaction among system elements in the forward direction
($\varphi^{i} + r^{i \crd{\alpha}}_{j} \varphi^{j}$)
with probability
$\tensor*[^{+}]{s}{_{\crd{\alpha}}}(\varphi^{k})$
and in the backward direction
($\varphi^{i} - r^{i \crd{\alpha}}_{j} \varphi^{j}$)
with probability
$\tensor*[^{-}]{s}{_{\crd{\alpha}}}(\varphi^{k})$
(see Fig.~\ref{fig:one-step_process}). In this case, the transition
probability matrix can be
written as
\begin{equation}
  w_{\crd{\alpha}}(\varphi^{i}| \psi^{i} ,t) = \tensor*[^{+}]{s}{_{\crd{\alpha}}}
  \delta_{\varphi^{i},\psi^{i}+1} + \tensor*[^{-}]{s}{_{\crd{\alpha}}}
  \delta_{\varphi^{i}, \psi^{i}-1},
  \qquad \crd{\alpha} = \overline{1,s},
\end{equation}
where $\delta_{i,j}$ is the Kronecker delta.

Thus, the general form of master equation~\eqref{eq:master:subansemble} for the state vector $\varphi^{i}$,
which is modified step by step with the step length $r^{i \crd{\alpha}}_j \varphi^j$, is as follows:
\begin{multline} 
  \label{eq:one-step:master:full}
  \pdv{p(\varphi^{i} ,t)}{t} = 
  \sum_{\crd{\alpha}=1}^{s} 
  \left\{
    \tensor*[^{-}]{s}{_{\crd{\alpha}}} (\varphi^{i}+r^{i
      \crd{\alpha}},t) p(\varphi^{i}+r^{i \crd{\alpha}} ,t) 
    + {} \right. \\ \left. {} + 
    \tensor*[^{+}]{s}{_{\crd{\alpha}}} (\varphi^{i}-r^{i \crd{\alpha}}, t) p(\varphi^{i} - r^{i \crd{\alpha}},t)
    - {} \right. \\ \left. {} - 
    % -
    \qty[ 
    \tensor*[^{+}]{s}{_{\crd{\alpha}}} (\varphi^{i}) + 
    \tensor*[^{-}]{s}{_{\crd{\alpha}}} (\varphi^{i})
    ]
    p(\varphi^{i},t)      
  \right\}.
\end{multline}

  The functions
  $\tensor*[^{+}]{s}{_{\crd{\alpha}}}$
  and
  $\tensor*[^{-}]{s}{_{\crd{\alpha}}}$
  for Eq.~\eqref{eq:one-step:master:full} are obtained using the combinatorial approach.

  The unit-time transition rates
  $\tensor*[^{+}]{s}{_{\crd{\alpha}}}$
  and
  $\tensor*[^{-}]{s}{_{\crd{\alpha}}}$
  are proportional to the number of
  ways for selecting the number of arrangements from
  $\varphi^{\crd{i}}$ to  $I^{\crd{i} \crd{\alpha}}$ (denoted as
  $A_{\varphi^{\crd{i}}}^{I^{\crd{i} \crd{\alpha}}}$)
  and from $\varphi^{\crd{i}}$ to $F^{\crd{i} \crd{\alpha}}$ (denoted
  as $A_{\varphi^{\crd{i}}}^{F^{\crd{i} \crd{\alpha}}}$),
  respectively; these rates are given by
  the expressions
\begin{equation}
  \label{eq:s-pm}
\begin{gathered}
  \tensor*[^{+}]{s}{_{\crd{\alpha}}} =
  \tensor*[^{+}]{k}{_{\crd{\alpha}}} \prod_{\crd{i}=1}^{n} 
  A_{\varphi^{\crd{i}}}^{I^{\crd{i} \crd{\alpha}}} =
  \tensor*[^{+}]{k}{_{\crd{\alpha}}} \prod_{\crd{i}=1}^{n}
  \frac{\varphi^{\crd{i}}!}{(\varphi^{\crd{i}} - I^{\crd{i} \crd{\alpha}})!}, \\
  \tensor*[^{-}]{s}{_{\crd{\alpha}}} = 
  \tensor*[^{-}]{k}{_{\crd{\alpha}}} \prod_{\crd{i}=1}^{n} 
  A_{\varphi^{\crd{i}}}^{F^{\crd{i} \crd{\alpha}}} =
  \tensor*[^{-}]{k}{_{\crd{\alpha}}} \prod_{\crd{i}=1}^{n}
  \frac{\varphi^{\crd{i}}!}{(\varphi^{\crd{i}}-F^{\crd{i} \crd{\alpha}})!}.
\end{gathered}
\end{equation}

\subsection{Fokker--Planck Equation}

The Fokker--Planck equation is a special case of the master equation
and can be regarded as its approximate form. It can be derived by
expanding the master equation in a series up to the second-order terms
inclusive. For this purpose, we can use the Kramers--Moyal expansion~\cite{gardiner:stochastic::en}
(for simplicity, it is written for a one-dimensional case):
\begin{equation}
  \label{eq:kramers-moyal}
  \pdv{p(\varphi,t)}{t} = \sum \limits_{n=1}^\infty \frac{(-1)^n}{n!}
  \pdv[n]{}{\varphi} \left[\xi^n(\varphi) p(\varphi,t)\right],
\end{equation}
where
\begin{equation}
  \xi^n(\varphi) = \int\limits_{-\infty}^{\infty} (\psi - \varphi)^{n}
  w(\psi|\varphi) \dd{\psi}.
\end{equation}

By dropping the terms of the order higher than the second, we obtain
the Fokker--Planck equation
\begin{equation}
  \label{eq:fp_1D}
  \pdv{p(\varphi, t)}{t} = 
  - \pdv{\varphi} \left[ A(\varphi) p(\varphi, t) \right] +
  \pdv[2]{}{\varphi} 
  \left[ B(\varphi) p(\varphi, t) \right],
\end{equation}
or, in a multidimensional case, 
\begin{multline}
  \label{eq:fp_nD}
  \pdv{p(\varphi^k, t)}{t} = 
  - \pdv{}{\varphi^{i}} \left[ A^{i}(\varphi^{k}) p(\varphi^{k}, t)
  \right] 
  + {} \\ {} +
  \frac{1}{2} \pdv{}{\varphi^{i}}{\varphi^{j}} 
  \left[ B^{i j} (\varphi^k) p(\varphi^{k}, t)  \right],
\end{multline}
where
\begin{equation} 
  \label{eq:fp_coeff}
  \begin{gathered}
    A^{i} := A^{i}(\varphi^{k}) = r^{i \crd{\alpha}} 
    \qty[
      \tensor*[^+_{\text{fp}}]{s}{_{\crd{\alpha}}} -
      \tensor*[^-_{\text{fp}}]{s}{_{\crd{\alpha}}} ], \\
    B^{i j} := B^{i j}(\varphi^{k}) = r^{i \crd{\alpha}} r^{j
      \crd{\alpha}} 
    \qty[ \tensor*[^+_{\text{fp}}]{s}{_{\crd{\alpha}}} -
      \tensor*[^-_{\text{fp}}]{s}{_{\crd{\alpha}}} ].
  \end{gathered}
\end{equation}

Using the Kramers--Moyal expansion, in~\eqref{eq:s-pm}, we can replace the
combinations of the form $\varphi (\varphi-1) \cdots (\varphi -
(n-1))$ by $(\varphi)^n$; as a result, for the Fokker--Planck
equation, we obtain
\begin{equation}
  \label{eq:s-pm:exp}
\begin{gathered}
  \tensor*[^{+}_{\text{fp}}]{s}{_{\crd{\alpha}}} =
  \tensor*[^{+}]{k}{_{\crd{\alpha}}} \prod_{\crd{i}=1}^{n}
  (\varphi^{\crd{i}})^{I^{\crd{i} \crd{\alpha}}}, \\
  \tensor*[^{-}_{\text{fp}}]{s}{_{\crd{\alpha}}} = 
  \tensor*[^{-}]{k}{_{\crd{\alpha}}} \prod_{\crd{i}=1}^{n}
  (\varphi^{\crd{i}})^{F^{\crd{i} \crd{\alpha}}}.
\end{gathered}
\end{equation}

It can be seen from~\eqref{eq:fp_coeff} that the coefficients of the Fokker--Planck
equation can be derived directly from~\eqref{eq:r_i} and~\eqref{eq:s-pm}, i.e., in this case,
there is no need to write the master equation.

\subsection{Langevin Equation}

  The Langevin equation 
\begin{equation}
  \label{eq:langevin}
  \dd \varphi^{i} = a^{i} \dd{t} + b^i_{a} \dd{W^{a}},
\end{equation}
  corresponds to the Fokker--Planck equation; here,
  $a^{i} := a^{i} (\varphi^k)$,
  $b^{i}_{a} := b^{i}_{a} (\varphi^k)$, $\varphi^i \in \setR^n $
  is the state
  vector of the system, and
  $W^{a} \in \setR^m$
  is the $m$-dimensional Wiener process. The
  Wiener process is implemented as
  $\dd{W} = \varepsilon \sqrt{\dd{t}}$,
  where
  $\varepsilon \sim N(0,1)$
  is the normal distribution
  with mean $0$ and variance $1$. Here, Latin indices from the middle of the
  alphabet denote the values associated with state vectors (space
  dimension is $n$), while Latin indices from the beginning of the alphabet
  denote the values associated with the vector of the Wiener process
  (space dimension is $m \leqslant n$).

  In this case, the coefficients of Eqs.~\eqref{eq:fp_nD} and~\eqref{eq:langevin} are related as
  follows:
\begin{equation}
  \label{eq:k-langevin}
  A^{i} = a^{i}, \qquad B^{i j} = b^{i}_{a} b^{j a}.
\end{equation}

It can be seen that the second term of the Langevin equation is a
square root, which has a complex form in the multidimensional
case. Note, however, that it is the squared second term of the
Langevin equation that is used in many relationships, which is why
there is usually no need to compute the root explicitly.

\section{Implementing the Model of One-Step Stochastic Processes in a Computer Algebra System}
\label{sec:cas}

\subsection{Justifying the Selection of a Computer Algebra System}
\label{sec:cas:compare}

To implement the algorithms considered above, we had to solve a
problem of selecting a computer algebra system. Our requirements quite
correspond to the requirements for a universal computer algebra
system; however, the spectrum of such systems is fairly wide. So, our
selection criteria were as follows:
\begin{itemize}
\item the system must be freeware (namely, open source system); this is our
  \emph{profession de foi};
\item the system must be interactive and support the read-–eval-–print
  loop (REPL) paradigm;
\item it is desirable that the system be supported and improved by a
  community of developers: it makes no sense to create a product in a
  language that may soon turn out to be dead;
\item symbolic computations constitute only one stage of the method,
  and the equations derived also need to be investigated, most often,
  by numerical methods; therefore, the system must support various
  output formats and, even better, provide seamless integration of a
  target system with other software products;
\item it is also desirable that the system enable implementation of
  numerical methods.
\end{itemize}

There are not so many universal open source computer algebra
systems. Let us consider several main candidates.

Maxima~\cite{bainov:2011:maxima,timberlake:2016:maxima} is a classical
system that, however, stopped in its
development at the level of the late 1990s. New versions are released
often but only to improve stability and correct errors. New
capabilities are introduced extremely slow. Moreover, the interaction
with other programs is limited.

Axiom~\cite{jenks:1992:axiom,ef-kor-gev-kul-sev:vestnik-miph:2014-3}
is distinguished for its mathematical approach to
computer algebra. This system supports the Hindley–Milner type
system~\cite{hindley:1969:type-scheme,milner:1978:type_polymorphism}
and has a powerful internal extension language. However,
because of the unresolved problems with copyright, the system is ``in
fever.'' Several versions of the system originated with each
modification having its own plan for development. It is still not
clear how and when this situation will end. Moreover, the
interoperability of this system is almost zero.

In our opinion, the most promising system is
SymPy~\cite{lamy:sympy_starter,kulyabov:2017:mpmm:diagram}. This
system was originally developed as a symbolic computation library for
Python. Quite unexpectedly, Python has become a universal glue
language. Its application in various projects caused an explosive
growth of related tools and libraries. And SymPy was developing
alongside it. Currently, SymPy is quite a powerful computer algebra
system. However, most of our criteria are satisfied not by the system
itself but by its libraries.

And yet, SymPy seems to satisfy all criteria listed above:
\begin{itemize}
\item as an interactive shell, it is convenient to use the Jupyter
  notebook (which is a component of the iPython
  system~\cite{perez:2007:ipython}) that supports the REPL paradigm;
\item Python is actually used as a glue language that allows one to
  integrate different software products, while the SciPy
  library~\cite{oliphant:2007:scipy} supports many output formats;
\item SymPy's output data can be naturally transferred for numerical
  computations to the NumPy library~\cite{oliphant:guide_numpy}.
\end{itemize}

Thus, to implement the method for stochastization of one-step
processes, we selected the SymPy system.

    \subsection{Software Implementation of the Stochastization Algorithm}
\label{sec:programm}

The algorithm for deriving a stochastic differential equation from an
interaction scheme (see algorithms~\ref{alg:stochast:full}
and~\ref{alg:stochast:simple}) is implemented as a
sequence of operations on vector data. The initial data (interaction
schemes) are represented as follows:
\begin{itemize}
\item a symbolic vector $X$ represents the state vector $\varphi$; 
\item a symbolic vector $K$ represents the interaction rates
  $\tensor*[^{+}]{k}{_{\crd{\alpha}}}$ and
  $\tensor*[^{-}]{k}{_{\crd{\alpha}}}$~\eqref{eq:chemkin}; 
\item numerical matrices $I$ and $F$ represent the initial and final states in
formula~\eqref{eq:n^i-notion}.
\end{itemize}

Main computations are carried out by four functions. 
The first function simply finds the element $\frac{\varphi^{\crd{i}}!}{(\varphi^{\crd{i}} - I^{\crd{i}
    \crd{\alpha}})!}$ from formula~\eqref{eq:s-pm}:
\begin{lstlisting}{language=python}
def P(x, n):
    """x is a symbol, n is an integer"""   
    return sp.prod([x-i for i in range(n)])
\end{lstlisting}

The second function uses the first one to compute
$\tensor*[^{+}]{s}{_{\crd{\alpha}}}$
and
$\tensor*[^{-}]{s}{_{\crd{\alpha}}}$;
this function
receives symbolic vectors
$X = (x^1,x^2,\ldots,x^n)^{T}$
and
$K = (\tensor*[^{-}]{k}{_{1}},\ldots, \tensor*[^{-}]{k}{_{s}})$
as its arguments and returns (as a
result) a list
$\tensor*[^{+}]{s}{_{1}},\ldots, \tensor*[^{+}]{s}{_{s}}$
(the code is presented only for forward reactions):
\begin{lstlisting}{language=python}
def S(X, K, I):
    res = []
    for i in range(len(K)):
        # Compute the elements of the product
        Ps = [P(x, int(n)) for (x, n) in zip(X, I[i, :])]
        # Find the product itself 
        res.append(K[i]* sp.prod(Ps))
    # Obtain the list [s_1, s_2, s_3, ..., s_s]
    return res
\end{lstlisting}

The following are the main functions of the algorithm:
\verb|drift_vector(X, K, I, F)| yields a drift vector and
\verb|diffusion_matrix(X, K, I, F)|
yields
a diffusion matrix (in the \emph{SymPy} symbolic format).
\begin{lstlisting}{language=python}
def drift_vector(X, K, I, F):
    """Drift vector"""
    res = sp.zeros(r=len(X), c=1)
    R = F.T - I.T
    for i in range(len(K)):
        res += R[:, i] * S(X, K, I)[i]
    return res
\end{lstlisting}

\begin{lstlisting}{language=python}
def diffusion_matrix(X, K, I, F):
    """Diffusion matrix"""
    res = sp.zeros(r=len(X), c=len(X))
    R = F.T - I.T
    R = sp.Matrix(R)
    for i in range(len(K)):
        res += R[:, i] * R[:, i].T * S(X, K, I)[i]
    return res
\end{lstlisting}

The output data of the program can be exported in the \LaTeX{} format by
using built-in methods that allow
$A^{i}$ and $B^{i j}$
to be translated into the
\LaTeX{} code. For this purpose, it is sufficient to call the combination
of functions \verb|print(sympy.latex(A))|, where the variable \verb|A| contains the
result of the function \verb|drift_vector| and the function \verb|latex()| exports
it into the \LaTeX{} code.

When using the \emph{Jupyter} interactive shell, the correct display of the
\TeX{} notation needs to be pretuned. For this purpose, it is required to
import the \texttt{Latex} module at the beginning of the \emph{Jupyter} notebook:
\begin{lstlisting}{language=python}
from IPython.display import Latex
\end{lstlisting}
and then call the function  
\begin{lstlisting}{language=python}
sympy.init_printing(use_unicode=True)
\end{lstlisting}

In addition, the expressions obtained can be exported in the format
supported by a wide spectrum of programming languages, including
universal (\emph{C/C++}, \emph{Fortran}) and domain-specific (\emph{Matematica}, \emph{Julia})
languages. These functions are available as methods of the
\verb|sympy.printing| class. For example, the following code yields a \emph{C++}
function \verb|vector<double> f(vector<double> x)|:
\begin{lstlisting}{language=python}
print("""vector<double> f(vector<double> x) {{
    vector<double> res = {{ {0} }};
    return res;
    }}""".format(", ".join([sp.printing.ccode(f) for f in f])))
\end{lstlisting}

\subsection{Implementation Example: Verhulst Model}
\label{sec:model:verhulst}

For demonstration of the method, let us consider the Verhulst
model~\cite{verhulst:1838,Feller:1939:acta_biotheoretica,feller:1949:theory_stochastic_processes}. In
population dynamics, this model describes limited population growth.

The deterministic model has the form   
\begin{equation}
  \label{eq:verhulst}
  \Dot{\varphi} = \lambda \varphi - \beta \varphi - \gamma \varphi^{2},
\end{equation}
where $\lambda$ is the reproduction rate, $\beta$ is the death rate,
and $\gamma$ is the population decline rate.

Based on~\eqref{eq:verhulst}, we write interaction scheme~\eqref{eq:chemkin2}:
\begin{equation}
  \label{eq:verhulst:schema}
  \begin{gathered}
    \varphi \overset{\lambda}{\underset{\gamma}{\rightleftharpoons}} 2
    \varphi ,
    \\
    0 \xleftarrow{\beta} \varphi.
  \end{gathered}
\end{equation}

Here, the first direct relationship~\eqref{eq:verhulst:schema} means
reproduction of an
individual, the first backward relationship means conflict between
individuals, and the second backward relationship means death of an
individual.

This model is one-dimensional ($n=1$). The number of interactions is $s = 2$. 

Based on~\eqref{eq:verhulst:schema}, we write the matrices
$I^{\crd{i}\crd{\alpha}}$
and
$F^{\crd{i}\crd{\alpha}}$:
\begin{equation}
  I^{\crd{i}\crd{\alpha}} = 
  \begin{pmatrix}
    1 \\
    1
  \end{pmatrix}
  , \quad
  F^{\crd{i}\crd{\alpha}} = 
  \begin{pmatrix}
    2\\
    0
  \end{pmatrix}
  .
\end{equation}

The following data are inputted to the program:
\begin{lstlisting}{language=python}
I = sp.Matrix([[1], [1]])
F = sp.Matrix([[2], [0]])
X = sp.Matrix(['phi'])
K = sp.Matrix(['k_{0}'.format(i+1) for i in range(2)])
\end{lstlisting}

Since the problem has one dimension, the drift vector and the
diffusion matrix are scalars:
\begin{equation}
  \begin{gathered}
    A(\varphi) =
    \lambda \varphi - \beta \varphi - \gamma \varphi^2
    ,\\
    B(\varphi) =
    \lambda \varphi + \beta \varphi - \gamma \varphi^2
    .
  \end{gathered}
\end{equation}

The stochastic differential equation corresponding to Eq.~\eqref{eq:verhulst} has the form 
\begin{equation}
  \dd{\varphi(t)} = (\lambda \varphi -\beta\varphi - \gamma
  \varphi^2)\dd{t} +
 \sqrt {(\lambda \varphi + \beta\varphi - \gamma \varphi^2)} \dd{W(t)}.
\end{equation}

Thus, our goal is achieved: the model is stochastized. It should be
noted that, for manual computations, even this simple one-dimensional
model is rather cumbersome.

\subsection{Implementation Example: Predator–Prey Model}
\label{sec:pp-model}

Systems that describe interaction between two types of population,
predators and preys, are well investigated, and there are many various
models for these systems. The model developed (independently of each
other) by
A.~Lotka~\cite{lotka:1910:periodic_reaction,lotka:1925:physical_biology}
and V.~Volterra~\cite{volterra:1928:fluctuations} is believed to be
the
first predator–prey model.

The deterministic system can be written as   
\begin{equation} 
  \label{eq:lotka-volterra} 
  \begin{dcases}
    \Dot{x} = k_1 x - k_{2} x y, \\
    \Dot{y} = k_{2} x y - k_3 y,
  \end{dcases}
\end{equation}
where
$x$ and $y$
are interacting individuals (prey and predator,
respectively) and $k_1,k_2,k_3$ are interaction rates.

The state vector is introduced as
$\varphi^{\crd{i}} = (x,y)^T$.
Based on~\eqref{eq:lotka-volterra}, we write interaction
scheme~\eqref{eq:chemkin2}:
\begin{equation} 
  \label{eq:lv-chemkin}
  \begin{gathered}
    x \xrightarrow{k_1} 2x,  \\
    x + y \xrightarrow{k_2} 2y,  \\
    y \xrightarrow{k_3} 0. 
  \end{gathered}
\end{equation}

Scheme~\eqref{eq:lv-chemkin} has the standard interpretation. The first relationship
means that the prey consumes a unit of food to immediately reproduce
itself. The second relationship means that the predator eats the prey
to immediately reproduce itself; this is the only possibility for prey
to die. The last relationship is the only possibility for predator to
die.

For this model, the system dimension is $n = 2$ and the number of interactions is $s = 3$.

Based on~\eqref{eq:lv-chemkin}, we write the matrices
$I^{\crd{i}\crd{\alpha}}$ and $F^{\crd{i}\crd{\alpha}}$:
\begin{equation}
  I^{\crd{i}\crd{\alpha}} = 
  \begin{pmatrix}
    1 & 0\\
    1 & 1\\
    0 & 1
  \end{pmatrix}
  , \quad
  F^{\crd{i}\crd{\alpha}} = 
  \begin{pmatrix}
    2 & 0\\
    0 & 2\\
    0 & 0
  \end{pmatrix}
  .
  % ,\;\;
  % \mathbf{R} = 
  % \begin{pmatrix}
  %    1 &  0\\
  %   -1 &  1\\
  %    0 & -1
  % \end{pmatrix}
\end{equation}

The following data are inputted to the program:
\begin{lstlisting}{language=python}
I = sp.Matrix([[1, 1, 0], [0, 1, 1]])
F = sp.Matrix([[2, 0, 0], [0, 2, 0]])
X = sp.Matrix(['x', 'y'])
K = sp.Matrix(['k_{0}'.format(i+1) for i in range(3)])
\end{lstlisting}

As a result, the program computes the drift vector and the diffusion
matrix:
\begin{equation}
  \begin{gathered}
    A^{\crd{i}}(x,y) =
    \begin{pmatrix}
      k_{1} x - k_{2} x y\\
      k_{2} x y - k_{3} y
    \end{pmatrix}
    ,\\
    B^{\crd{i} \crd{j}}(x,y) =
    \begin{pmatrix}
      k_{1} x + k_{2} x y & - k_{2} x y \\
      - k_{2} x y & k_{2} x y + k_{3} y
    \end{pmatrix}
    .
  \end{gathered}
\end{equation}

These relationships allow us to write the Fokker--Planck equation and
the stochastic differential equation corresponding to it. For this
purpose, it is sufficient to extract the square root of the matrix
$B^{\crd{i} \crd{j}}$:

\begin{equation}
  \label{eq:pp_langevin}
  \begin{gathered}
    \dd 
    \begin{pmatrix}
      x \\
      y
    \end{pmatrix}
    =
    \begin{pmatrix}
      k_1 x-k_2xy\\
      k_2xy - k_3y
    \end{pmatrix}
    \dd{t} + 
    b^{\crd{i}}_{\crd{a}} 
    \begin{pmatrix}
      \dd{W^{1}} \\
      \dd{W^{2}}
    \end{pmatrix},\\
    b^{\crd{i}}_{a} b^{\crd{j} a} = B^{\crd{i}\crd{j}} =
    \begin{pmatrix}
      k_1 x+k_2xy & -k_2xy \\
      -k_2xy & k_2xy + k_3y
    \end{pmatrix}.
  \end{gathered}
\end{equation}

It can be seen that the deterministic part of system~\eqref{eq:pp_langevin} coincides
with the original system~\eqref{eq:lotka-volterra}. Since the root of a matrix in an
analytical form is quite difficult to extract, it would be optimal to
investigate this system numerically.

\section{Conclusion}
\label{sec:conclusion}

Research process often has an iterative character. Computational
results are estimated based on certain criteria. Computations have to
be carried out repeatedly. Computer algebra systems make it possible
to automate this process.

In this paper, we have considered the simplest implementation of the
algorithm for stochastization of one-step processes by using
interaction schemes. For this purpose, we have analyzed several
computer algebra systems based on the proposed criteria. According to
these criteria, as a computer algebra system for implementation of the
stochastization method, we have selected the \emph{SymPy} system. Some
essential code fragments of the implemented stochastization method
have been presented. The operation of the software complex has been
demonstrated on the Verhulst and Lotka--Volterra models.

For all its seeming simplicity, this software product significantly
improves labor productivity of the researcher. Thus, it can be
concluded that computer algebra systems, in addition to numerical
computation systems, have become a necessary tool for the researcher.

\begin{acknowledgments}

The publication has been prepared with the support of the ``RUDN University Program 5-100'' 
and funded by Russian Foundation for Basic Research (RFBR) according to the research project 
No~16-07-00556.

\end{acknowledgments}

 \bibliographystyle{elsarticle-num}

\bibliography{bib/sdu-generation/cite}

\end{document}

\graphicspath{{image/}}

\title{Реализация метода стохастизации одношаговых процессов в системе
компьютерной алгебры}

\author{М. Н. Геворкян}
\email{gevorkyan_mn@rudn.university}
\affiliation{Кафедра прикладной информатики и теории вероятностей,\\
  Российский университет дружбы народов,\\
  ул. Миклухо-Маклая, д. 6, Москва, Россия, 117198}

\author{А. В. Демидова}
\email{demidova_av@rudn.university}
\affiliation{Кафедра прикладной информатики и теории вероятностей,\\
  Российский университет дружбы народов,\\
  ул. Миклухо-Маклая, д. 6, Москва, Россия, 117198}

\author{Т. Р. Велиева}
\email{trvelieva@gmail.com}
\affiliation{Кафедра прикладной информатики и теории вероятностей,\\
  Российский университет дружбы народов,\\
  ул. Миклухо-Маклая, д. 6, Москва, Россия, 117198}

\author{А. В. Королькова}
\email{korolkova_av@rudn.university}
\affiliation{Кафедра прикладной информатики и теории вероятностей,\\
  Российский университет дружбы народов,\\
  ул. Миклухо-Маклая, д. 6, Москва, Россия, 117198}

\author{Д. С. Кулябов}
\email{kulyabov_ds@rudn.university}
\affiliation{Кафедра прикладной информатики и теории вероятностей,\\
  Российский университет дружбы народов,\\
  ул. Миклухо-Маклая, д. 6, Москва, Россия, 117198}
\affiliation{Лаборатория информационных технологий,\\
  Объединённый институт ядерных исследований,\\
  ул. Жолио-Кюри 6, Дубна, Московская область, Россия, 141980}

\author{Л. А. Севастьянов}
\email{sevastianov_la@rudn.university}
\affiliation{Кафедра прикладной информатики и теории вероятностей,\\
  Российский университет дружбы народов,\\
  ул. Миклухо-Маклая, д. 6, Москва, Россия, 117198}
\affiliation{Лаборатория теоретической физики,\\
  Объединённый институт ядерных исследований,\\
  ул. Жолио-Кюри 6, Дубна, Московская область, Россия, 141980}

\begin{abstract}
  При моделировании таких явлений как популяционная динамика,
  исследование управляемых потоков и т.д. возникает проблема адаптации
  существующих моделей под исследуемое явление. Для этого
  предлагается получать новые модели из первых принципов на основе
  метода стохастизации одношаговых процессов..
  Исследование имеет вид итеративного процесса, заключающегося в
  получении модели и последующей её корректировке. Количество таких
  итераций может быть крайне большим.
  Целью данной работы является разработка программной реализации
  средствами компьютерной алгебры метода стохастизации одношаговых
  процессов.
  В работе предложено использовать систему компьютерной алгебры
  \emph{SymPy} в качестве основы для программной реализации. На основе
  разработанного алгоритма показано получение стохастических
  дифференциальных уравнений их вида схем взаимодействия.
  Результаты работы программы продемонстрированы на моделях Ферхюльста
  и Лотки--Вольтерры.
\end{abstract}

  % \keywords{активное управление трафиком, теория управления,
  %   автоколебательный режим}

\maketitle

\SetKwInput{KwData}{Исходные параметры}
\SetKwInput{KwResult}{Результат}
\SetKwInput{KwIn}{Входные данные}
\SetKwInput{KwOut}{Выходные данные}
\SetKwIF{If}{ElseIf}{Else}{если}{тогда}{иначе если}{иначе}{конец условия}
\SetKwFor{While}{до тех пор, пока}{выполнять}{конец цикла}
\SetKw{KwTo}{от}
\SetKw{KwRet}{возвратить}
\SetKw{Return}{возвратить}
\SetKwBlock{Begin}{начало блока}{конец блока}
\SetKwSwitch{Switch}{Case}{Other}{Проверить значение}{и выполнить}{вариант}{в противном случае}{конец варианта}{конец проверки значений}
\SetKwFor{For}{цикл}{выполнять}{конец цикла}
\SetKwFor{ForEach}{для каждого}{выполнять}{конец цикла}
\SetKwRepeat{Repeat}{повторять}{до тех пор, пока}
\SetAlgorithmName{Алгоритм}{алгоритм}{Список алгоритмов}

\section{Введение}

  Многие физические и технические явления можно описывать в рамках
  статистического подхода. Обычно подбирается модель, достаточно полно
  отражающая изучаемое явление, и в неё вносятся некоторые
  уточнения. Возникает вопрос, как вносить изменения, поскольку данный
  процесс не однозначен.
  Используемые модели являются реализацией некоторых первых
  принципов. Мы считаем, что если строить модели из первых принципов,
  то вносимые изменения будут выражать внутреннюю структуру модели.
  Для описания изучаемых нами явлений (сети передачи данных, системы с
  управлением, популяционная динамика) мы используем модель
  одношаговых
  процессов~\cite{gardiner:stochastic::ru,van-kampen:stochastic::ru}.

  Нами разработана методика стохастизации моделей, которая позволяет
  получать из первых принципов и стохастическую модель, и
  соответствующую ей
  детерминистическую~\cite{kulyabov:2016:ecms:one-step,kulyabov:2016:dccn-springer:diagram,kulyabov:2015:conf:mmcp-2015:doi,kulaybov:2017:ecms:wind,kulyabov:2016:rk-stochastic}.
  Процес исследования является итеративным: из детерминистической
  модели мы получаем первичную модель, из первичной модели
  стохастическую, стохастическая модель соотносится с
  детерминистической, на основании этого соотнесения мы получает
  уточнённую первичную модель. Далее процесс повторяется.
 
  Формализм стохастизации может быть реализован разными способами. На
  данный момент нами применяются представление векторов состояния
  (комбинаторный подход) и представление чисел заполнения (операторный
  подход)~\cite{grassberger:1980:fock-space, tauber:2005,
    janssen:2004, mobilia:2005}.

  В случае применения комбинаторного подхода все действия выполняются
  в пространстве векторов состояния системы. На протяжении всех
  модельных манипуляций мы имеем дело с конкретной исследуемой
  системой. В результате получается описание в виде дифференциального
  уравнения. Этот подход удобен при конструировании модели, поскольку
  позволяет легко сравнить результат с другими моделями.

  В операторном подходе мы отвлекаемся от конкретной реализации
  исследуемой системы, работая с абстрактными операторами. В
  пространство векторов состояний мы переходим только в конце
  вычислений. Кроме того, конкретную операторную алгебру мы выбираем,
  исходя их симметрии задачи.  Этот подход удобен при теоретических
  построениях.

  Зачастую возникает задача нахождения стохастической модели,
  эквивалентной ранее созданной детерминистической.  Для этого мы
  используем комбинаторный подход. В этом представлении стохастическая
  модель имеет вид дифференциального уравнения, что облегчает
  сравнение с исходной моделью.

Структура статьи следующая. В разделе~\ref{sec:notation} введены
основные обозначения и соглашения. Идеология метода стохастизации
одношаговых процессов и его
составные части описаны в разделе~\ref{sec:method}.
В разделе~\ref{sec:cas} описан программный комплекс, реализующий метод
стохастизации одношаговых процессов. В разделе~\ref{sec:cas:compare}
проводится сравнение систем компьютерной алгебры и выбор конкретной
системы для реализации. В разделе~\ref{sec:programm} описаны основные
фрагменты программы
реализации метода. В разделах~\ref{sec:model:verhulst}
и~\ref{sec:pp-model} приведены примеры конкретного применения метода.

\section{Обозначения и соглашения}
\label{sec:notation}

\begin{enumerate}

\item В работе для тензорных величин используются абстрактные
  индексы~\cite{penrose-rindler:spinors::ru}, то есть тензор как
  целостный объект обозначается просто индексом (например, $x^{i}$),
  а компоненты тензора обозначаются подчёркнутым индексом (например,
  $x^{\crd{i}}$).

\item Также будем придерживаться следующих соглашений. Латинские индексы из
  середины алфавита ($i$, $j$, $k$) будут относиться к пространству
  векторов состояний системы. Латинские индексы из начала алфавита
  ($a$) будут относиться к пространству винеровского
  процесса. Греческие индексы ($\alpha$) будут задавать количество
  разных взаимодействий в кинетических уравнениях.

\end{enumerate}

\section{Методика стохастизации детерминированных моделей}
\label{sec:method}

  Нами разработана эмпирическая методика стохастизации одношаговых
  процессов. Методика формализована таким образом, что для её применения
  достаточно сформулировать исходную задачу соответствующим образом.
  При этом лишь часть шагов явно алгоритмизирована. 

\begin{figure}
  \centering
  \includegraphics[width=\linewidth]{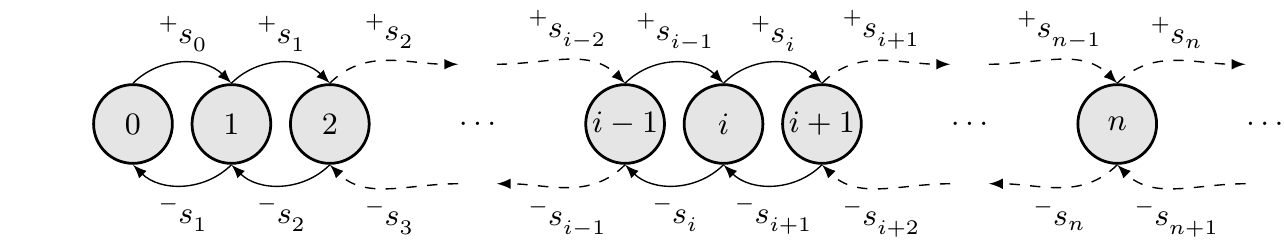}
  \caption{Одношаговый процесс}
  \label{fig:one-step_process}
\end{figure}

  Первым шагом мы приводим нашу модель к виду одношагового
  процесса~(см. рис.~\ref{fig:one-step_process}).  Далее необходимо
  формализовать этот процесс в виде схем
  взаимодействия~\cite{kulyabov:2014:icumt-2014:p2p,kulyabov:2015:conf:mmcp-2015:doi}.
  Аналогами схем взаимодействия являются уравнения химической
  кинетики, реакции частиц и т.д.

  Для схем взаимодействия создан алгоритм (вернее будет сказать,
  семейство алгоритмов --- для разных подходов разные алгоритмы),
  позволяющий из схем взаимодействия записать основное
  кинетическое уравнение. Однако это
  уравнение~\cite{van-kampen:stochastic::ru, gardiner:stochastic::ru}
  имеет обычно достаточно сложную структуру, что затрудняет его
  исследование и решение.  Следующим шагом мы получаем приближённые
  модели в виде уравнений Фоккера--Планка и Ланжевена.

  Предложенный подход подразумевает итеративность исследования:
  полученные приближённые модели уточняются, что приводит
  к коррекции исходных схем взаимодействия.

При построении моделей мы работаем по следующему итеративному алгоритму
(алгоритм~\ref{alg:stochast:real}).

\begin{algorithm}
  \caption{Алгоритм работы исследователя}
  \label{alg:stochast:real}
  \SetAlgoLined
  \KwData{детерминистическое уравнение}
  \KwResult{стохастическое уравнение}
  \While{не получено соответствие уравнений}{
    схема взаимодействия
    $\leftarrow$
    детерминистическое уравнение \;
    стохастическое уравнение
    $\leftarrow$
    схема взаимодействия\;
    \eIf{стохастическое уравнение удовлетворяет модели}{
      завершить стохастизацию\;
    }{
      вернуться к началу\;
    }
  }
\end{algorithm}

К сожалению, большая часть этого
алгоритма не формализована. Получение схем взаимодействия, определение
соответствия уравнений~--- всё это тяжело формализуемые задачи. Однако
проблема формализации получения стохастических уравнений из схем
взаимодействия нами решена (алгоритмы~\ref{alg:stochast:full}
и~\ref{alg:stochast:simple}).

\begin{algorithm}
  \SetAlgoLined
  \KwData{схема взаимодействия~\eqref{eq:chemkin}}
  \KwResult{уравнение Ланжевена~\eqref{eq:langevin} и~\eqref{eq:k-langevin}}
  \Begin{%
    \emph{Операторы состояния системы}\\
    $I_j^{i \alpha}$, $F_j^{i \alpha}$  $\leftarrow$ схема
    взаимодействия $\eqref{eq:chemkin}$\;
    \emph{Оператор изменения состояния системы}\\
    $r_j^{i \alpha}$ $\leftarrow$
    $I_j^{i \alpha}$, $F_j^{i \alpha}$ $\eqref{eq:r_i}$\;
    \emph{Интенсивности переходов}\\
    $\tensor*[^{+}]{s}{_{\crd{\alpha}}}$,
    $\tensor*[^{-}]{s}{_{\crd{\alpha}}}$ $\leftarrow$
    $I_j^{i \alpha}$, $F_j^{i \alpha}$, $r_j^{i \alpha}$ $\eqref{eq:s-pm}$\;
    основное кинетическое уравнение
    $\leftarrow$
    $\tensor*[^{+}]{s}{_{\crd{\alpha}}}$,
    $\tensor*[^{-}]{s}{_{\crd{\alpha}}}$,
    $r_j^{i \alpha}$
    $\eqref{eq:one-step:master:full}$ \;
    уравнение Фоккера--Планка
    $\leftarrow$
    основное кинетическое уравнение $\eqref{eq:kramers-moyal}$, $\eqref{eq:fp_nD}$ \;
    уравнение Ланжевена
    $\leftarrow$ уравнение Фоккера--Планка
    $\eqref{eq:langevin}$, $\eqref{eq:k-langevin}$ \;
  }
  \caption{Алгоритм стохастизации}
  \label{alg:stochast:full}
\end{algorithm}

\begin{algorithm}
  \SetAlgoLined
  \KwData{схема взаимодействия~\eqref{eq:chemkin}}
  \KwResult{уравнение Ланжевена~\eqref{eq:langevin} и~\eqref{eq:k-langevin}}
  \Begin{%
    \emph{Операторы состояния системы}\\
    $I_j^{i \alpha}$, $F_j^{i \alpha}$  $\leftarrow$ Схемa
    взаимодействия $\eqref{eq:chemkin}$\;
    \emph{Оператор изменения состояния системы}\\
    $r_j^{i \alpha}$ $\leftarrow$
    $I_j^{i \alpha}$, $F_j^{i \alpha}$ $\eqref{eq:r_i}$\;
    \emph{Интенсивности переходов}\\
    $\tensor*[^{+}_{\text{fp}}]{s}{_{\crd{\alpha}}}$,
    $\tensor*[^{-}_{\text{fp}}]{s}{_{\crd{\alpha}}}$ $\leftarrow$
    $I_j^{i \alpha}$, $F_j^{i \alpha}$, $r_j^{i \alpha}$ $\eqref{eq:s-pm:exp}$\;
    уравнение Ланжевена
    $\leftarrow$
    $\tensor*[^{+}_{\text{fp}}]{s}{_{\crd{\alpha}}}$,
    $\tensor*[^{-}_{\text{fp}}]{s}{_{\crd{\alpha}}}$,
    $r_j^{i \alpha}$
    $\eqref{eq:langevin}$, $\eqref{eq:k-langevin}$ \;
  }
  \caption{Упрощённый алгоритм стохастизации}
  \label{alg:stochast:simple}
\end{algorithm}

Процесс этот несложный, но крайне трудоёмкий. Опишем основные этапы этого процесса более подробно.

\subsection{Схемы взаимодействия}
\label{sec:schema}

Состояние системы будем описывать вектором состояний
$\varphi^{i} \in \setR^n$, где $n$~--- размерность
системы.
Оператор $I^{i}_{j} \in \setN^{n}_{0} \times \setN^{n}_{0}$ задаёт
состояние системы до взаимодействия, оператор
$F^{i}_{j} \in \setN^{n}_{0} \times \setN^{n}_{0}$~--- после.
Компонентные индексы размерности системы пробегают значения
$\crd{i},\crd{j} = \overline{1,n}$. В результате взаимодействия
происходит переход системы в другое состояние.

В системе может происходить $s$ видов различных
взаимодействий. Поэтому вместо операторов $I^{i}_{j}$ и $F^{i}_{j}$
будем рассматривать операторы
$I^{i \alpha}_{j} \in \setN^{n}_{0} \times \setN^{n}_{0} \times
\setN^{s}_{+}$
и
$F^{i \alpha}_{j} \in \setN^{n}_{0} \times \setN^{n}_{0} \times
\setN^{s}_{+}$.
Компонентные индексы количества взаимодействий пробегают значения
  $\crd{\alpha} = \overline{1,s}$.

Взаимодействие элементов системы описывается с помощью схем
взаимодействия.
\begin{equation}
  \label{eq:chemkin}
  I^{i \crd{\alpha}}_{j} \varphi^j
  \xrightleftharpoons[{\tensor*[^{-}]{k}{_{\crd{\alpha}}}}]{\tensor*[^{+}]{k}{_{\crd{\alpha}}}}
  F^{i \crd{\alpha}}_{j} \varphi^{j},
  \qquad \crd{\alpha} = \overline{1,s}.
\end{equation}

  Здесь греческие индексы задают количество взаимодействий, а
  латинские~--- размерность системы. Коэффициенты
  $\tensor*[^{+}]{k}{_{\crd{\alpha}}}$ и
  $\tensor*[^{-}]{k}{_{\crd{\alpha}}}$ имеют смысл интенсивности
  (скорости) взаимодействия.

Изменение состояния будет задаваться оператором
\begin{equation}
  \label{eq:r_i}
  r_j^{i \crd{\alpha}} = F_j^{i \crd{\alpha}} -I_j^{i \crd{\alpha}}.
\end{equation}

Таким образом, один шаг взаимодействия $\crd{\alpha}$ в прямом и
обратном направлениях можно записать соответственно как
\begin{equation}
  \begin{gathered}
    \varphi^{i}  \rightarrow \varphi^i + r^{i \crd{\alpha}}_{j} \varphi^{j},\\
    \varphi^{i} \rightarrow \varphi^{i} - r^{i \crd{\alpha}}_{j} \varphi^{j}.
  \end{gathered}
\end{equation}

  Чаще всего модель строится таким образом, что тензоры
  $I^{i \crd{\alpha}}_{j}$ и $F^{i \crd{\alpha}}_{j}$
  диагональны по
латинским индексам. Поэтому мы можем явно использовать диагональные
сечения этих матриц и 
записывать~\eqref{eq:chemkin} не в виде векторных уравнений, 
а в виде более традиционных сумм~\cite{van-kampen:stochastic::ru}:
\begin{equation}
  \label{eq:chemkin2}
  I^{i \crd{\alpha}}_{j} \varphi^j \delta_i
  \xrightleftharpoons[{\tensor*[^{-}]{k}{_{\crd{\alpha}}}}]{\tensor*[^{+}]{k}{_{\crd{\alpha}}}}
  F^{i \crd{\alpha}}_{j} \varphi^{j} \delta_i,
\end{equation}
где $\delta_{\crd{i}} = (1,\ldots,1)$.

Также мы будем использовать следующие обозначения:
\begin{equation}
  \label{eq:n^i-notion}
  I^{i \crd{\alpha}} := I^{i \crd{\alpha}}_{j} \delta^{j}, 
  \quad F^{i \crd{\alpha}} := F^{i \crd{\alpha}}_{j} \delta^{j}, \quad
  r^{i \crd{\alpha}} := r^{i \crd{\alpha}}_{j} \delta^{j}.
\end{equation}

\subsection{Основное кинетическое уравнение}
\label{sec:master-equation}

Для одношаговых процессов 
в качестве кинетического уравнения рассматривается основное
кинетическое уравнение~\cite{van-kampen:stochastic::ru,
  gardiner:stochastic::ru}. 
\begin{multline}
  \label{eq:master:trans}
  \pdv{p(\varphi_{2},t_{2}|\varphi_{1},t_{1})}{t} = \int \bigl[
  w(\varphi_{2}|\psi,t_{2}) p(\psi,t_{2}|\varphi_{1},t_{1}) 
  - {} \\ {} -
  w(\psi|\varphi_{2},t_{2}) p(\varphi_{2},t_{2}|\varphi_{1},t_{1}) 
  \bigr] \dd{\psi},
\end{multline}
где $w(\varphi|\psi,t)$~--- вероятность перехода из состояния $\psi$
в состояние $\varphi$ за единицу времени.

Основное кинетическое уравнение можно рассматривать как реализацию
уравнения Колмогорова. Однако основное кинетическое уравнение более
удобно и имеет непосредственную физическую
интерпретацию~\cite{van-kampen:stochastic::ru}.

  Зафиксировав начальные значения $\varphi_{1},t_{1}$, можно записать
  данное уравнение для подансамбля:
\begin{equation}
  \label{eq:master:subansemble}
  \pdv{p(\varphi,t)}{t} = \int
  \qty[
  w(\varphi|\psi,t) p(\psi,t) -
  w(\psi|\varphi,t) p(\varphi,t)
  ] \dd{\psi}.
\end{equation}

При дискретной области определения множества состояний системы $\varphi$ можно
записать~\eqref{eq:master:subansemble} (пронумеровав состояния числами
$n$ и $m$):
\begin{equation} 
  \label{eq:mas_eq}
    \pdv{p_{n}(t)}{t} = \sum\limits_{m} 
    \qty[w_{nm} p_{m}(t) - w_{mn} p_{n}(t)],
\end{equation}
где $p_{n}$~--- вероятность нахождения
системы в состоянии $n$ в момент времени $t$,
$w_{nm}$~--- вероятность перехода системы из состояния $m$ в
состояние $n$ за единицу времени.

  В системе, описываемой одношаговыми процессами, возможны два вида
  перехода системы из одного состояния в другое, происходящие в
  результате взаимодействия элементов в прямом направлении
  ($\varphi^{i} + r^{i \crd{\alpha}}_{j} \varphi^{j}$) с вероятностью
  $\tensor*[^{+}]{s}{_{\crd{\alpha}}}(\varphi^{k})$ и в обратном
  направлении ($\varphi^{i} - r^{i \crd{\alpha}}_{j} \varphi^{j}$) с
  вероятностью
  $\tensor*[^{-}]{s}{_{\crd{\alpha}}}(\varphi^{k})$~(рис.~\ref{fig:one-step_process}).
  А матрица вероятностей переходов может быть записана в виде:
\begin{equation}
  w_{\crd{\alpha}}(\varphi^{i}| \psi^{i} ,t) = \tensor*[^{+}]{s}{_{\crd{\alpha}}}
  \delta_{\varphi^{i},\psi^{i}+1} + \tensor*[^{-}]{s}{_{\crd{\alpha}}}
  \delta_{\varphi^{i}, \psi^{i}-1},
  \qquad \crd{\alpha} = \overline{1,s},
\end{equation}
где $\delta_{i,j}$ --- символ Кронекера.

  Таким образом, общий вид основного кинетического
  уравнения~\eqref{eq:master:subansemble} для вектора состояний
  $\varphi^{i}$, изменяющегося шагами длины
  $r^{i \crd{\alpha}}_j \varphi^j$, принимает вид:
\begin{multline} 
  \label{eq:one-step:master:full}
  \pdv{p(\varphi^{i} ,t)}{t} = 
  \sum_{\crd{\alpha}=1}^{s} 
  \left\{
    \tensor*[^{-}]{s}{_{\crd{\alpha}}} (\varphi^{i}+r^{i
      \crd{\alpha}},t) p(\varphi^{i}+r^{i \crd{\alpha}} ,t) 
    + {} \right. \\ \left. {} + 
    \tensor*[^{+}]{s}{_{\crd{\alpha}}} (\varphi^{i}-r^{i \crd{\alpha}}, t) p(\varphi^{i} - r^{i \crd{\alpha}},t)
    - {} \right. \\ \left. {} - 
    % -
    \qty[ 
    \tensor*[^{+}]{s}{_{\crd{\alpha}}} (\varphi^{i}) + 
    \tensor*[^{-}]{s}{_{\crd{\alpha}}} (\varphi^{i})
    ]
    p(\varphi^{i},t)      
  \right\}.
\end{multline}

  Получим функции $\tensor*[^{+}]{s}{_{\crd{\alpha}}}$ и
  $\tensor*[^{-}]{s}{_{\crd{\alpha}}}$ для
  уравнения~\eqref{eq:one-step:master:full}, исходя из комбинаторного
  подхода.

  Интенсивности перехода в единицу времени
  $\tensor*[^{+}]{s}{_{\crd{\alpha}}}$ и
  $\tensor*[^{-}]{s}{_{\crd{\alpha}}}$ пропорциональны соответственно
  числу способов выбора числа размещений из $\varphi^{\crd{i}}$ по
  $I^{\crd{i} \crd{\alpha}}$ (обозначается как
  $A_{\varphi^{\crd{i}}}^{I^{\crd{i} \crd{\alpha}}}$) и по
  $F^{\crd{i} \crd{\alpha}}$ (обозначается как
  $A_{\varphi^{\crd{i}}}^{F^{\crd{i} \crd{\alpha}}}$) и определяются
  выражениями:
\begin{equation}
  \label{eq:s-pm}
\begin{gathered}
  \tensor*[^{+}]{s}{_{\crd{\alpha}}} =
  \tensor*[^{+}]{k}{_{\crd{\alpha}}} \prod_{\crd{i}=1}^{n} 
  A_{\varphi^{\crd{i}}}^{I^{\crd{i} \crd{\alpha}}} =
  \tensor*[^{+}]{k}{_{\crd{\alpha}}} \prod_{\crd{i}=1}^{n}
  \frac{\varphi^{\crd{i}}!}{(\varphi^{\crd{i}} - I^{\crd{i} \crd{\alpha}})!}, \\
  \tensor*[^{-}]{s}{_{\crd{\alpha}}} = 
  \tensor*[^{-}]{k}{_{\crd{\alpha}}} \prod_{\crd{i}=1}^{n} 
  A_{\varphi^{\crd{i}}}^{F^{\crd{i} \crd{\alpha}}} =
  \tensor*[^{-}]{k}{_{\crd{\alpha}}} \prod_{\crd{i}=1}^{n}
  \frac{\varphi^{\crd{i}}!}{(\varphi^{\crd{i}}-F^{\crd{i} \crd{\alpha}})!}.
\end{gathered}
\end{equation}

\subsection{Уравнение Фоккера--Планка}

  Уравнение Фоккера--Планка является частным случаем основного
  кинетического уравнения и может рассматриваться как его приближённая
  форма. Его можно получить путём разложения основного кинетического
  уравнения в ряд до членов второго порядка включительно.  Для этого
  можно использовать разложение
  Крамерса--Мойала~\cite{gardiner:stochastic::ru} (для простоты
  записано в одномерном случае):
\begin{equation}
  \label{eq:kramers-moyal}
  \pdv{p(\varphi,t)}{t} = \sum \limits_{n=1}^\infty \frac{(-1)^n}{n!}
  \pdv[n]{}{\varphi} \left[\xi^n(\varphi) p(\varphi,t)\right],
\end{equation}
где
\begin{equation}
  \xi^n(\varphi) = \int\limits_{-\infty}^{\infty} (\psi - \varphi)^{n}
  w(\psi|\varphi) \dd{\psi}.
\end{equation}

  Отбрасывая члены выше второго порядка, получаем уравнение
  Фоккера--Планка:
\begin{equation}
  \label{eq:fp_1D}
  \pdv{p(\varphi, t)}{t} = 
  - \pdv{\varphi} \left[ A(\varphi) p(\varphi, t) \right] +
  \pdv[2]{}{\varphi} 
  \left[ B(\varphi) p(\varphi, t) \right],
\end{equation}
или в многомерном случае:
\begin{multline}
  \label{eq:fp_nD}
  \pdv{p(\varphi^k, t)}{t} = 
  - \pdv{}{\varphi^{i}} \left[ A^{i}(\varphi^{k}) p(\varphi^{k}, t)
  \right] 
  + {} \\ {} +
  \frac{1}{2} \pdv{}{\varphi^{i}}{\varphi^{j}} 
  \left[ B^{i j} (\varphi^k) p(\varphi^{k}, t)  \right],
\end{multline}
где
\begin{equation} 
  \label{eq:fp_coeff}
  \begin{gathered}
    A^{i} := A^{i}(\varphi^{k}) = r^{i \crd{\alpha}} 
    \qty[
      \tensor*[^+_{\text{fp}}]{s}{_{\crd{\alpha}}} -
      \tensor*[^-_{\text{fp}}]{s}{_{\crd{\alpha}}} ], \\
    B^{i j} := B^{i j}(\varphi^{k}) = r^{i \crd{\alpha}} r^{j
      \crd{\alpha}} 
    \qty[ \tensor*[^+_{\text{fp}}]{s}{_{\crd{\alpha}}} -
      \tensor*[^-_{\text{fp}}]{s}{_{\crd{\alpha}}} ].
  \end{gathered}
\end{equation}

  В результате разложения Крамерса--Мойала можно 
  в~\eqref{eq:s-pm} заменить комбинации типа
  $\varphi (\varphi-1) \cdots (\varphi - (n-1))$ на $(\varphi)^n$, 
  получим для уравнения Фоккера--Планка:
\begin{equation}
  \label{eq:s-pm:exp}
\begin{gathered}
  \tensor*[^{+}_{\text{fp}}]{s}{_{\crd{\alpha}}} =
  \tensor*[^{+}]{k}{_{\crd{\alpha}}} \prod_{\crd{i}=1}^{n}
  (\varphi^{\crd{i}})^{I^{\crd{i} \crd{\alpha}}}, \\
  \tensor*[^{-}_{\text{fp}}]{s}{_{\crd{\alpha}}} = 
  \tensor*[^{-}]{k}{_{\crd{\alpha}}} \prod_{\crd{i}=1}^{n}
  (\varphi^{\crd{i}})^{F^{\crd{i} \crd{\alpha}}}.
\end{gathered}
\end{equation}

  Как видно из~\eqref{eq:fp_coeff}, коэффициенты уравнения
  Фоккера--Планка можно получить сразу из~\eqref{eq:r_i}
  и~\eqref{eq:s-pm}, то есть в данном случае записывать основное
    кинетическое уравнение нет необходимости.

\subsection{Уравнение Ланжевена}

  Уравнению Фоккера--Планка соответствует уравнение Ланжевена:
\begin{equation}
  \label{eq:langevin}
  \dd \varphi^{i} = a^{i} \dd{t} + b^i_{a} \dd{W^{a}},
\end{equation}
  где $a^{i} := a^{i} (\varphi^k)$,
  $b^{i}_{a} := b^{i}_{a} (\varphi^k)$, $\varphi^i \in \setR^n $ ---
  вектор состояний системы, $W^{a} \in \setR^m$ --- $m$-мерный
  винеровский процесс.  Винеровский процесс реализуется как
  $\dd{W} = \varepsilon \sqrt{\dd{t}}$, где
  $\varepsilon \sim N(0,1)$~--- нормальное распределение со средним
  $0$ и дисперсией $1$. Здесь латинскими индексами из середины
  алфавита обозначаются величины, относящиеся к векторам состояний
  (размерность пространства $n$), а латинскими индексами из начала
  алфавита обозначаются величины, относящиеся к вектору винеровского
  процесса (размерность пространства $m \leqslant n$).

  При этом связь между коэффициентами уравнений~\eqref{eq:fp_nD}
  и~\eqref{eq:langevin} выражается следующими соотношениями:
\begin{equation}
  \label{eq:k-langevin}
  A^{i} = a^{i}, \qquad B^{i j} = b^{i}_{a} b^{j a}.
\end{equation}

  Видно, что второй член уравнения Ланжевена представляет из себя
  квадратный корень, который имеет сложный вид в многомерном
  случае. Впрочем заметим, что во многих соотношениях используется именно квадрат
  второго члена уравнения Ланжевена, так что явное вычисление корня зачастую
  и не требуется.

\section{Реализация модели одношаговых стохастических процессов в системе компьютерной алгебры}
\label{sec:cas}

\subsection{Обоснование выбора системы компьютерной алгебры}
\label{sec:cas:compare}

При реализации рассмотренных алгоритмов перед нами встала задача
выбора системы компьютерной алгебры. Наши потребности вполне
укладываются в требования к универсальной системе компьютерной
алгебры,  но спектр таких систем % универсальных систем компьютерной алгебры
весьма
широк. Поэтому приведём наши критерии выбора:
\begin{itemize}
\item Система должна быть свободно распространяемой. Можно считать это
  нашим \emph{profession de foi}.
\item Взаимодействие с системой должно быть итеративным. В ней должна
  быть реализована парадигма REPL (Read--Eval--Print Loop).
\item Желательно, чтобы система поддерживалась и
  развивалось некоторым сообществом разработчиков. Неприятно создавать
  продукт на языке, который в скором
  времени окажется мёртвым.
\item Символьные вычисления --- лишь один из этапов
  методики. Полученные уравнения необходимо исследовать чаще всего
  численными методами. Поэтому необходимо иметь возможность разных форматов
  вывода результатов. А ещё лучше, иметь возможность бесшовно
  интегрировать искомую систему с другими программными продуктами. 
\item Также было бы удобно иметь реализацию численных методов в рамках
  системы компьютерной алгебры.
\end{itemize}

Выбор универсальных свободных систем компьютерной алгебры не так уж и
велик. Рассмотрим несколько основных претендентов.

Система Maxima~\cite{bainov:2011:maxima,timberlake:2016:maxima}~---
классическая система, однако застывшая в своём развитии в конце
90-х. Новые версии выпускаются часто. При этом они лишь увеличивают
стабильность, исправляют ошибки. Добавление новых возможностей идёт
крайне медленно.  Кроме того, возможность взаимодействия с другими
программами ограничена.

Система
Axiom~\cite{jenks:1992:axiom,ef-kor-gev-kul-sev:vestnik-miph:2014-3}
выделяется математическим подходом к компьютерной алгебре. Она
поддерживает систему типов
Хиндли--Милнера~\cite{hindley:1969:type-scheme,milner:1978:type_polymorphism},
обладает 
великолепным внутренним языком расширения. Но из-за неразрешённых
проблем с копирайтом систему лихорадит. Образовалось несколько
форков. Каждый вариант имеет свою идеологию, свои планы развития. Чем
и когда всё это закончится~--- не понятно. Да и интероперабельность
фактически отсутствует.

Наиболее интересной для нас является система
SymPy~\cite{lamy:sympy_starter,kulyabov:2017:mpmm:diagram}. Эта
система появилась как библиотека символьных вычислений для языка
Python. Но язык Python стал универсальным языковым клеем (достаточно
неожиданно). Применение его в разнообразных проектах привело к
взрывному росту сопутствующих средств и библиотек. Поэтому и SymPy
развивался вместе с ним. Теперь это достаточно мощная система
компьютерной алгебры. Причём большая часть необходимых нам критериев
проистекают не из собственно системы SymPy, а из окружающих её библиотек.

Получается, что SymPy
удовлетворяет всем нашим критериям:
\begin{itemize}
\item В качестве интерактивной оболочки удобно использовать блокнот
Jupyter, являющийся компонентом системы
iPython~\cite{perez:2007:ipython}, реализующей идеологию REPL.
\item Язык Python фактически используется как соединительный язык,
  своего рода язык-клей, который позволяет интегрировать между собой
  разные программные продукты. Кроме того, в рамках библиотеки
  SciPy~\cite{oliphant:2007:scipy} поддерживается большое число
  выходных форматов.
\item Выходные данные SymPy возможно естественным образом передать для
  численных расчётов в библиотеку NumPy~\cite{oliphant:guide_numpy}.
\end{itemize}

Таким образом, мы остановились в своём выборе на системе SymPy для
реализации метода стохастизации одношаговых процессов.

  \subsection{Программная реализации алгоритма стохастизации}
\label{sec:programm}

Алгоритм получения стохастического дифференциального уравнения из
схемы взаимодействия (алгоритмы~\ref{alg:stochast:full} и~\ref{alg:stochast:simple}) реализован
как последовательность операций над векторными данными. Исходными
данными являются схемы взаимодействия, представленные в следующем
виде:
\begin{itemize}
\item символьный вектор $X$ представляет собой вектор состояния
  системы $\varphi$;
\item символьный вектор $K$ представляет собой интенсивности
  взаимодействия $\tensor*[^{+}]{k}{_{\crd{\alpha}}}$,
  $\tensor*[^{-}]{k}{_{\crd{\alpha}}}$~\eqref{eq:chemkin};
\item числовые матрицы $I$ и $F$ представляют собой начальное и
  конечное состояния~\eqref{eq:n^i-notion}.
\end{itemize}

Основные вычисления реализованы в виде четырёх функций.

Первая функция просто реализует нахождение элемента
$\frac{\varphi^{\crd{i}}!}{(\varphi^{\crd{i}} - I^{\crd{i}
    \crd{\alpha}})!}$~\eqref{eq:s-pm}:
\begin{lstlisting}{language=python}
def P(x, n):
    """x is a symbol, n is an integer"""   
    return sp.prod([x-i for i in range(n)])
\end{lstlisting}

Следующая функция использует предыдущую для вычисления
$\tensor*[^{+}]{s}{_{\crd{\alpha}}}$ и
$\tensor*[^{-}]{s}{_{\crd{\alpha}}}$, в качестве аргументов
здесь %функции
передаются символьные векторы $X = (x^1,x^2,\ldots,x^n)^{T}$ и
$K = (\tensor*[^{-}]{k}{_{1}},\ldots, \tensor*[^{-}]{k}{_{s}})$, а в
качестве результата % Функция возвращает
возвращается список
$\tensor*[^{+}]{s}{_{1}},\ldots, \tensor*[^{+}]{s}{_{s}}$ (приведён
код только для прямых реакций):
\begin{lstlisting}{language=python}
def S(X, K, I):
    res = []
    for i in range(len(K)):
        # Compute the elements of the product
        Ps = [P(x, int(n)) for (x, n) in zip(X, I[i, :])]
        # Find the product itself 
        res.append(K[i]* sp.prod(Ps))
    # Obtain the list [s_1, s_2, s_3, ..., s_s]
    return res
\end{lstlisting}

Следующие функции~--- основные функции алгоритма: для получения вектора сноса
\verb|drift_vector(X, K, I, F)| и матрицы диффузии
\verb|diffusion_matrix(X, K, I, F)| в символьном формате
\emph{SymPy}:
\begin{lstlisting}{language=python}
def drift_vector(X, K, I, F):
    """Drift vector"""
    res = sp.zeros(r=len(X), c=1)
    R = F.T - I.T
    for i in range(len(K)):
        res += R[:, i] * S(X, K, I)[i]
    return res
\end{lstlisting}

\begin{lstlisting}{language=python}
def diffusion_matrix(X, K, I, F):
    """Diffusion matrix"""
    res = sp.zeros(r=len(X), c=len(X))
    R = F.T - I.T
    R = sp.Matrix(R)
    for i in range(len(K)):
        res += R[:, i] * R[:, i].T * S(X, K, I)[i]
    return res
\end{lstlisting}

Результаты работы программы можно экспортировать в формат \LaTeX{}.
Можно воспользоваться встроенными
методами, которые позволяют преобразовать $A^{i}$ и $B^{i j}$
в  код \LaTeX{}. Для этого достаточно вызвать комбинацию функций
\verb|print(sympy.latex(A))|, где в переменной \verb|A| содержится
результат работы функции \verb|drift_vector|, а функция \verb|latex()|
экспортирует его в \LaTeX{}-код.

При использовании интерактивной оболочки \emph{Jupyter} необходимо предварительно
настроить корректное отображение нотации \TeX{}.  Для этого вначале
\emph{Jupyter}-блокнота следует импортировать модуль \texttt{Latex}:
\begin{lstlisting}{language=python}
from IPython.display import Latex
\end{lstlisting}
и вызвать функцию
\begin{lstlisting}{language=python}
sympy.init_printing(use_unicode=True)
\end{lstlisting}

Кроме того, можно экспортировать получившиеся выражения в формат,
соответствующий синтаксису большого спектра языков
программирования. Поддерживаются как универсальные языки
программирования (\emph{C/C++}, \emph{Fortran}), так и проблемно-специфичные
(\emph{Matematica}, \emph{Julia}). Данные функции доступны как методы класса
\verb|sympy.printing|. Например, результатом работы следующего кода
будет \emph{C++}-функция \verb|vector<double> f(vector<double> x)|:
\begin{lstlisting}{language=python}
print("""vector<double> f(vector<double> x) {{
    vector<double> res = {{ {0} }};
    return res;
    }}""".format(", ".join([sp.printing.ccode(f) for f in f])))
\end{lstlisting}

\subsection{Пример реализации. Модель Ферхюльста}
\label{sec:model:verhulst}

  Для демонстрации метода рассмотрим модель
  Ферхюльста~\cite{verhulst:1838,Feller:1939:acta_biotheoretica,feller:1949:theory_stochastic_processes}.
  В популяционной динамике эта модель описывает ограниченный рост
  популяции.

  Детерминистическая модель имеет следующий вид:
\begin{equation}
  \label{eq:verhulst}
  \Dot{\varphi} = \lambda \varphi - \beta \varphi - \gamma \varphi^{2},
\end{equation}
где $\lambda$~--- коэффициент интенсивности размножения, $\beta$~---
коэффициент интенсивности вымирания, $\gamma$~--- коэффициент
интенсивности уменьшения популяции.

  На основании~\eqref{eq:verhulst} запишем схему
  взаимодействия~\eqref{eq:chemkin2}:
\begin{equation}
  \label{eq:verhulst:schema}
  \begin{gathered}
    \varphi \overset{\lambda}{\underset{\gamma}{\rightleftharpoons}} 2
    \varphi ,
    \\
    0 \xleftarrow{\beta} \varphi.
  \end{gathered}
\end{equation}

  Первое прямое соотношение~\eqref{eq:verhulst:schema} означает
  репродукцию особи, первое обратное соотношение~-- соперничество между
  особями. Второе соотношение описывает гибель особи.

  Данная модель одномерна ($n=1$). Количество взаимодействий $s$ равно~2.

  Из~\eqref{eq:verhulst:schema} запишем матрицы
  $I^{\crd{i}\crd{\alpha}}$ и $F^{\crd{i}\crd{\alpha}}$:
\begin{equation}
  I^{\crd{i}\crd{\alpha}} = 
  \begin{pmatrix}
    1 \\
    1
  \end{pmatrix}
  , \quad
  F^{\crd{i}\crd{\alpha}} = 
  \begin{pmatrix}
    2\\
    0
  \end{pmatrix}
  .
\end{equation}

В программу передаются следующие значения:
\begin{lstlisting}{language=python}
I = sp.Matrix([[1], [1]])
F = sp.Matrix([[2], [0]])
X = sp.Matrix(['phi'])
K = sp.Matrix(['k_{0}'.format(i+1) for i in range(2)])
\end{lstlisting}

Поскольку задача одномерна, то 
вектор сноса и матрица диффузии являются скалярами:
\begin{equation}
  \begin{gathered}
    A(\varphi) =
    \lambda \varphi - \beta \varphi - \gamma \varphi^2
    ,\\
    B(\varphi) =
    \lambda \varphi + \beta \varphi - \gamma \varphi^2
    .
  \end{gathered}
\end{equation}

Стохастическое дифференциальное уравнение,
соответствующее уравнению~\eqref{eq:verhulst}, будет иметь вид:
\begin{equation}
  \dd{\varphi(t)} = (\lambda \varphi -\beta\varphi - \gamma
  \varphi^2)\dd{t} +
 \sqrt {(\lambda \varphi + \beta\varphi - \gamma \varphi^2)} \dd{W(t)}.
\end{equation}

Таким образом мы достигли своей цели --- модель
стохастизирована. Следует заметить, что даже такая простая одномерная
модель при ручных расчётах является достаточно трудоёмкой.

\subsection{Пример реализации. Модель <<хищник--жертва>>}
\label{sec:pp-model}

Системы, описывающие взаимодействие двух видов популяций типа <<хищник--жертва>>,
широко исследованы, и для таких систем существует большое количество
разнообразных моделей. Самой первой моделью <<хищник--жертва>> принято
считать модель, полученную независимо друг от друга А.~Лоткой~\cite{lotka:1910:periodic_reaction,lotka:1925:physical_biology} и
В.~Вольтеррой~\cite{volterra:1928:fluctuations}.

Детерминистическую систему можно записать в следующем виде:
\begin{equation} 
  \label{eq:lotka-volterra} 
  \begin{dcases}
    \Dot{x} = k_1 x - k_{2} x y, \\
    \Dot{y} = k_{2} x y - k_3 y,
  \end{dcases}
\end{equation}
где $x$, $y$ --- взаимодействующие сущности ($x$~--- жертва,
$y$~--- хищник), коэффициенты $k_1,k_2,k_3$~---
интенсивности взаимодействия.

  Введём вектор состояний $\varphi^{\crd{i}} = (x,y)^T$.  На
  основании~\eqref{eq:lotka-volterra} запишем схему
  взаимодействия~\eqref{eq:chemkin2}:
\begin{equation} 
  \label{eq:lv-chemkin}
  \begin{gathered}
    x \xrightarrow{k_1} 2x,  \\
    x + y \xrightarrow{k_2} 2y,  \\
    y \xrightarrow{k_3} 0. 
  \end{gathered}
\end{equation}

  Схема~\eqref{eq:lv-chemkin} имеет стандартную интерпретацию.  Первое
  соотношение означает, что жертва, которая съедает единицу пищи,
  немедленно репродуцируется. Второе соотношение описывает поглощение
  жертвы хищником и мгновенное репродуцирование хищника. Это
  единственная рассматриваемая возможность гибели жертвы. Последнее
  соотношение --- это естественная смерть хищника.

Для данной модели размерность системы $n$ равна 2, количество
взаимодействий $s$ равно 3.

Из~\eqref{eq:lv-chemkin} запишем 
матрицы $I^{\crd{i}\crd{\alpha}}$ и $F^{\crd{i}\crd{\alpha}}$:
\begin{equation}
  I^{\crd{i}\crd{\alpha}} = 
  \begin{pmatrix}
    1 & 0\\
    1 & 1\\
    0 & 1
  \end{pmatrix}
  , \quad
  F^{\crd{i}\crd{\alpha}} = 
  \begin{pmatrix}
    2 & 0\\
    0 & 2\\
    0 & 0
  \end{pmatrix}
  .
  % ,\;\;
  % \mathbf{R} = 
  % \begin{pmatrix}
  %    1 &  0\\
  %   -1 &  1\\
  %    0 & -1
  % \end{pmatrix}
\end{equation}

В программу передаются следующие значения:
\begin{lstlisting}{language=python}
I = sp.Matrix([[1, 1, 0], [0, 1, 1]])
F = sp.Matrix([[2, 0, 0], [0, 2, 0]])
X = sp.Matrix(['x', 'y'])
K = sp.Matrix(['k_{0}'.format(i+1) for i in range(3)])
\end{lstlisting}

В результате будут вычислены вектор сноса и матрица диффузии
\begin{equation}
  \begin{gathered}
    A^{\crd{i}}(x,y) =
    \begin{pmatrix}
      k_{1} x - k_{2} x y\\
      k_{2} x y - k_{3} y
    \end{pmatrix}
    ,\\
    B^{\crd{i} \crd{j}}(x,y) =
    \begin{pmatrix}
      k_{1} x + k_{2} x y & - k_{2} x y \\
      - k_{2} x y & k_{2} x y + k_{3} y
    \end{pmatrix}
    .
  \end{gathered}
\end{equation}

  Эти соотношения позволяют записать непосредственно уравнение
  Фоккера--Планка и соответствующее ему стохастическое
  дифференциальное уравнение.  Для этого достаточно извлечь квадратный
  корень из матрицы $B^{\crd{i} \crd{j}}$:

\begin{equation}
  \label{eq:pp_langevin}
  \begin{gathered}
    \dd 
    \begin{pmatrix}
      x \\
      y
    \end{pmatrix}
    =
    \begin{pmatrix}
      k_1 x-k_2xy\\
      k_2xy - k_3y
    \end{pmatrix}
    \dd{t} + 
    b^{\crd{i}}_{\crd{a}} 
    \begin{pmatrix}
      \dd{W^{1}} \\
      \dd{W^{2}}
    \end{pmatrix},\\
    b^{\crd{i}}_{a} b^{\crd{j} a} = B^{\crd{i}\crd{j}} =
    \begin{pmatrix}
      k_1 x+k_2xy & -k_2xy \\
      -k_2xy & k_2xy + k_3y
    \end{pmatrix}.
  \end{gathered}
\end{equation}

Отсюда видно, что детерминистическая часть
системы~\eqref{eq:pp_langevin} совпадает с исходной
системой~\eqref{eq:lotka-volterra}.  Поскольку операция извлечения
корня из матрицы в аналитическом виде представляет из себя достаточно
сложную задачу, исследование этой системы оптимальнее проводить
численно.

\section{Заключение}
\label{sec:conclusion}

  Исследовательская работа часто носит итеративный
  характер. Результаты вычислений оцениваются по определённым
  критериям. Эти действия приходится выполнять многократно.
  Системы компьютерной алгебры позволяют автоматизировать этот процесс.
  
  В данной работе рассмотрена простейшая реализация алгоритма
  стохастизации одношаговых процессов по заданным схемам
  взаимодействия.  Для этого был проведён анализ систем компьютерной
  алгебры на основе предложенных авторами критериев. В соответствии с
  этими критериями в качестве системы компьютерной алгебры для
  реализации метода выбрана система \emph{SymPy}.
  Приведены существенные элементы кода реализации метода стохастизации
  одношаговых процессов.
  Работа программного комплекса продемонстрирована на примере двух
  моделей: модели Ферхюльста и модели Лотки--Вольтерры.

  При всей своей кажущейся простоте данный программный продукт резко
  увеличивает производительность труда исследователя. Можно сделать
  вывод, что системы компьютерной алгебры, наряду с системами
  численных расчётов, стали необходимым
  инструментом исследователя.

\def\acknowledgmentsname{Благодарности}

\begin{acknowledgments}

Работа частично поддержана грантами РФФИ \textnumero~15-07-08795, 16-07-00556.
Также публикация подготовлена при поддержке программы РУДН <<5-100>>.

\end{acknowledgments}

  \ifthenelse{\boolean{xetex}}
  {\bibliographystyle{ugost2008l}}
  {\bibliographystyle{gost2008l}}

\bibliography{bib/sdu-generation/cite}